\input{aipcheck}
\documentclass[
    ,final            % use final for the camera ready runs
  ]
  {aipproc}

\layoutstyle{6x9}

\def\Journal#1#2#3#4{{#1} {\bf#2}, {#3} {(#4)}}
\def\NPA{{ Nucl. Phys. A}}
\def\PLB{{ Phys. Lett.}  B}

\def\PPNP{{ Prog. Part. Nucl. Phys.}}
\def\PRP{{ Phys. Rep.}}
\def\PRL{ Phys. Rev. Lett.}

\def\PRD{{ Phys. Rev.} D}
\def\PRC{{ Phys. Rev.} C}

\def\ZPC{{Z. Phys.} C}

\def\EPJC{{Eur. Phys. J.} C}

\def\MPLA{{Mod. Phys. Lett.} A}

\def\JHEP{J. High Energy Phys.}

\def\ra{\rightarrow}

\def\be{\begin{equation}}
\def\ee{\end{equation}}
\def\bea{\begin{eqnarray}}
\def\eea{\end{eqnarray}}

\def\qbar{{\bar q}}
\def\ubar{{\bar u}}
\def\dbar{{\bar d}}
\def\sbar{{\bar s}}

\def\PPNP{{Prog. Part. Nucl. Phys.}}
\def\NPB{{ Nucl. Phys. B}}
\def\NPA{{ Nucl. Phys. A}}

\begin{document}

\title{Applications of Symmetry Breaking in
Determining PDFs of the Nucleon}

\classification{14.20.Dh, 13.88+e., 12.39.Ba}
\keywords      {Symmetry Breaking, Parton Distribution Function, Meson Cloud Model}

\author{Fu-Guang Cao}{
  address={Institute of Fundamental Sciences, Massey University, Privative Bag 11 222, Palmerston North, New Zealand}}

\begin{abstract}

Studying the possible breaking of various parton model symmetries by the parton
distribution functions of the nucleon can provide important information for the non-perturbative structure
of hadrons and the strong interaction. We review theoretical calculations for 
the breaking of flavor symmetry, quark-antiquark symmetry and charge symmetry
in the unpolarized and polarized nucleons using the meson cloud model.
We report an estimation for the total distribution of strange and antistrange quarks in the nucleon
by  combining theoretical calculations of SU(3) flavor symmetry breaking 
with light antiquark distributions obtained from global analysis of available experimental data.
 
\end{abstract}

\maketitle

%%%%%%%%%%%%%%%%%%%%%%%%%%%%%%%%%%%%%%%%%%%%
%% MAINMATTER
%%%%%%%%%%%%%%%%%%%%%%%%%%%%%%%%%%%%%%%%%%%%

\section{Introduction}

The possible breaking of parton model symmetries by the nucleon's parton 
distribution functions (PDFs) has been a topic of great interest due to its sensitivity to the non-perturbative dynamics of the nucleon.
In particular, the flavor asymmetry in the unpolarized nucleon sea
has been confirmed by several experiments \cite{NA51,E866,HERMES98}, which has impacted the determination
of both valence and sea quark distributions of the nucleon~\cite{JCPeng01}.

There are two dominant mechanisms for the quark sea production in the nucleon: (I) gluons splitting into quak-antiquark pairs,
and (II) non-pertutbative processes such as nucleons fluctuating into virtual meson-baryon pairs with the partons in the baryon and 
meson acting as the sea quarks in the nucleon.
While the sea distributions generated
through mechanism (I) can be assumed to be flavor independent (SU(3) flavor symmetric), i.e. $\dbar=\ubar=\sbar$ and
$d_{sea}=u_{sea}=s_{sea}$ and quark-antiquark symmetric, i.e. $\qbar=q$, the sea distributions generated through mechanism (II)
violate these symmetries. Mechanism (II) provides a natural explanation for the
observed SU(2) flavor asymmetry among the sea distributions ($\dbar > \ubar$) since 
 the probability of the Fock state $|n\pi^+\rangle$  is larger than that of the
$|\Delta^{++}\pi^-\rangle$ state in the proton wave function \cite{Thomas83,MST98,Kumano}. Calculations using mechanism (II) also
predict strange-antistrange symmetry breaking \cite{SignalT87,FCaoS_ssbar99,FCaoS_KKstar}
and charge symmetry breaking \cite{FCaoS_CSB} by the PDFs of the nucleon,
which may play an important role in explaining NuTeV's anomalous results \cite{NuTeV02}  for the $\sin^{2} \theta_{W}$ extracted 
from neutrino deep inelastic scattering (DIS) processes \cite{Davidson02}.

The total distribution of the strange and antistrange quarks ($s + \sbar$) 
in the nucleon is not well determined compared with those for the light quark sea.
The experimental constraints for the $s+\sbar$ come mainly from CCFR \cite{CCFR} and NuTeV \cite{NuTeV}  measurements of 
neutrino(antineutrino)-nucleon DIS processes, 
and HERMES measurement of charge kaon production in deep inelastic scattering on the deuteron \cite{HERMES08}.
A precise understanding on the cross-section for $W$ production at the Large Hadron Collider (LHC)
depends on the strange sea distributions at small $x$ region.

In this paper, we report investigations on the breaking of flavor symmetry \cite{FCaoS_plzedEPJC,FCaoS_plzedPRD},
quark-antiquark symmetry \cite{FCaoS_ssbar99,FCaoS_KKstar} and charge symmetry \cite{FCaoS_CSB}
by the nucleon's PDFs in the unpolarized and polarized nucleons using the meson cloud model.
Combining theoretical calculations of SU(3) flavor symmetry breaking 
with light antiquark distributions obtained from global analysis of available experimental data we can estimate
the total distribution of strange and antistrange quarks in the nucleon.

\section{Meson Cloud Model}

The meson cloud model (MCM) \cite{Thomas83,Kumano} is a model of non-perturbative contributions to the
parton distributions of the nucleon. The essential point of the MCM is that the nucleon can fluctuate into different
baryon-meson Fock states, $p \ra B M$.
The model assumes that the lifetime of a virtual baryon-meson Fock state is much
longer than the interaction time in the deep inelastic or Drell-Yan
process, thus the quark and antiquark in the virtual baryon-meson Fock states
can contribute to the parton distributions of the nucleon.

The contributions to the unpolarized quark and antiquark distributions of the nucleon can be written as,
\bea
 \delta q(x)&=&\sum_{BM} \left[ \int^1_x \frac{dy}{y}  f_{BM/N} (y) q^{B}(\frac{x}{y})
 + \int^1_x \frac{dy}{y}  f_{MB/N} (y) q^{M}(\frac{x}{y}) \right], \label{xq} \\
 \delta\qbar (x)&=&\sum_{BM} \int^1_x \frac{dy}{y}  f_{MB/N} (y)
\qbar^M(\frac{x}{y}), \label{xqbar}
\eea
where $f_{BM}(y)=f_{MB}(1-y)$ are unpolarized fluctuation functions giving the probability of finding the baryon with 
longitudinal momentum fraction $y$ and the meson with $1-y$ in the Fock state $|BM \rangle$,
\bea
f_{BM/N} (y)&=&\sum_{\lambda \lambda^\prime}
\int^\infty_0 d k_\perp^2
\phi^{\lambda \lambda^\prime}_ {B M}(y, k_\perp^2)
\phi^{*\,\lambda \lambda^\prime}_{B M}(y, k_\perp^2).
\label{eq:ffBM}
\eea
The wave function $\phi^{\lambda \lambda^\prime}_ {B M}(y, k_\perp^2)$ for the Fock state containing a baryon ($B$) with longitudinal 
momentum $y$, transverse momentum ${\bf k}_\perp$, and helicity $\lambda$, and a meson with momentum fraction $y$,
transverse momentum $-{\bf k}_\perp$, and helicity $\lambda^\prime$
can be derived from effective meson-nucleon Lagrangians
employing time-order perturbation theory in the infinite momentum frame \cite{HHoltmannSS}.
For the contributions to the polarized quark and antiquark distributions we refer the readers to \cite{FCaoS_plzedEPJC,FCaoS_plzedPRD}.
The parton distribution functions for the baryon and meson ($q^B$, $q^M$ and $\qbar^M$) can be calculated using the MIT bag model \cite{Bag_Adelaide}.

\section{Symmetry Breaking in the Parton Distribution Functions of the Nucleon}

As the contributions from the baryon-meson Fock states are not as constrained by the quark level symmetries of the
nucleon wave function, they may break various symmetries predicted by the naive parton model.

\subsection{Flavour symmetry breaking in the polarized nucleon sea}

The SU(2) flavor symmetry breaking in the unpolarized nucleon sea is well established, both experimentally \cite{NA51,E866,HERMES98}
and theoretically \cite{MST98,Kumano}.
We calculated the SU(2) flavor symmetry breaking in the polarized nucleon sea by
taking into account fluctuations $p \ra N \pi$, $N \rho$, $N \omega$, $\Delta \pi$ and $\Delta \rho$ \cite{FCaoS_plzedEPJC,FCaoS_plzedPRD},
\bea  
\Delta \dbar -\Delta \ubar =
\int^1_x \frac{dy}{y} \left[ \frac{2}{3} \Delta f _{\rho N/N} (y) - \frac{1}{3}\Delta f_{\rho \Delta/N}(y) \right]
\Delta V_\rho (\frac{x}{y}),
\label{eq:Ddubar}
\eea
where $\Delta f _{\rho N/N} (y)$ and $\Delta f_{\rho \Delta/N}(y)$ are polarized fluctuation functions and
$\Delta V_\rho (x)$ is the polarized valence parton distribution of the $\rho$ meson.
The contributions from possible interferences between different baryon-meson Fock states are much smaller than that given by Eq.~(\ref{eq:Ddubar}).

\begin{figure}[htb]
\includegraphics[width=14pc]{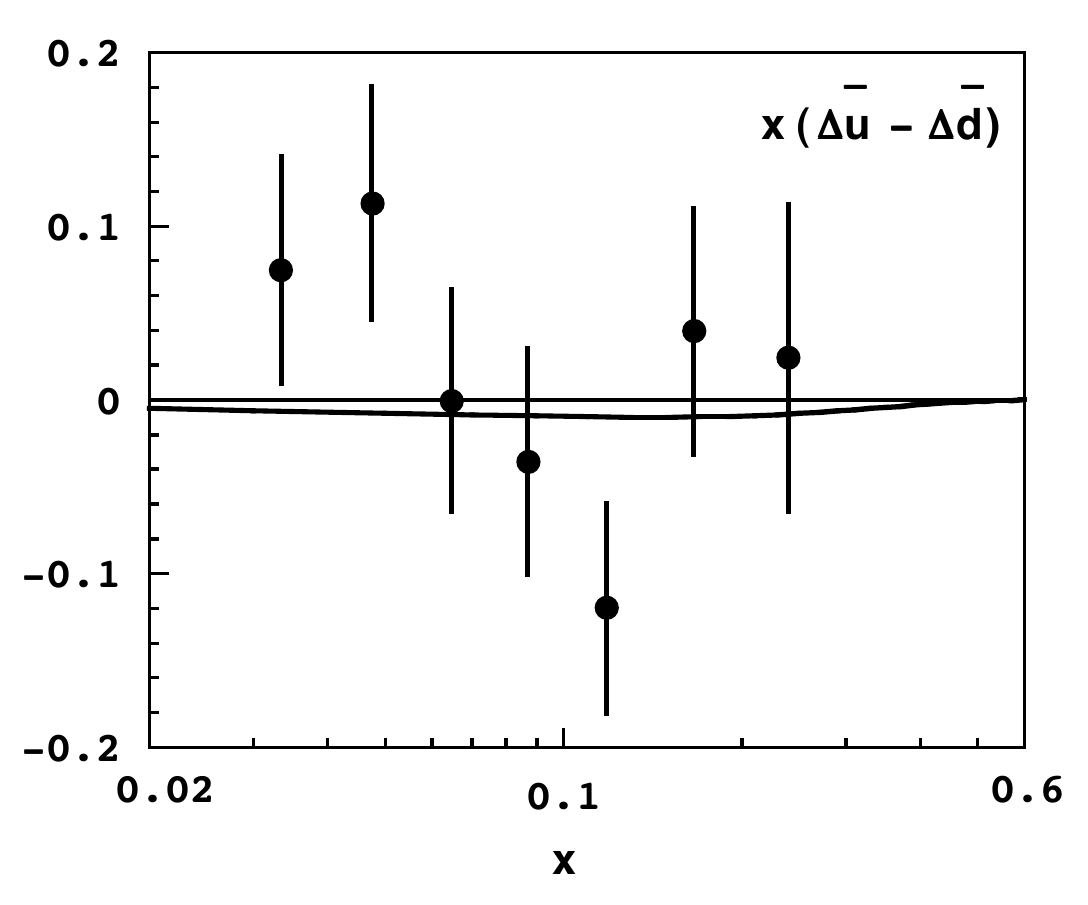} {\hskip 0.5cm}
\includegraphics[width=18.5pc]{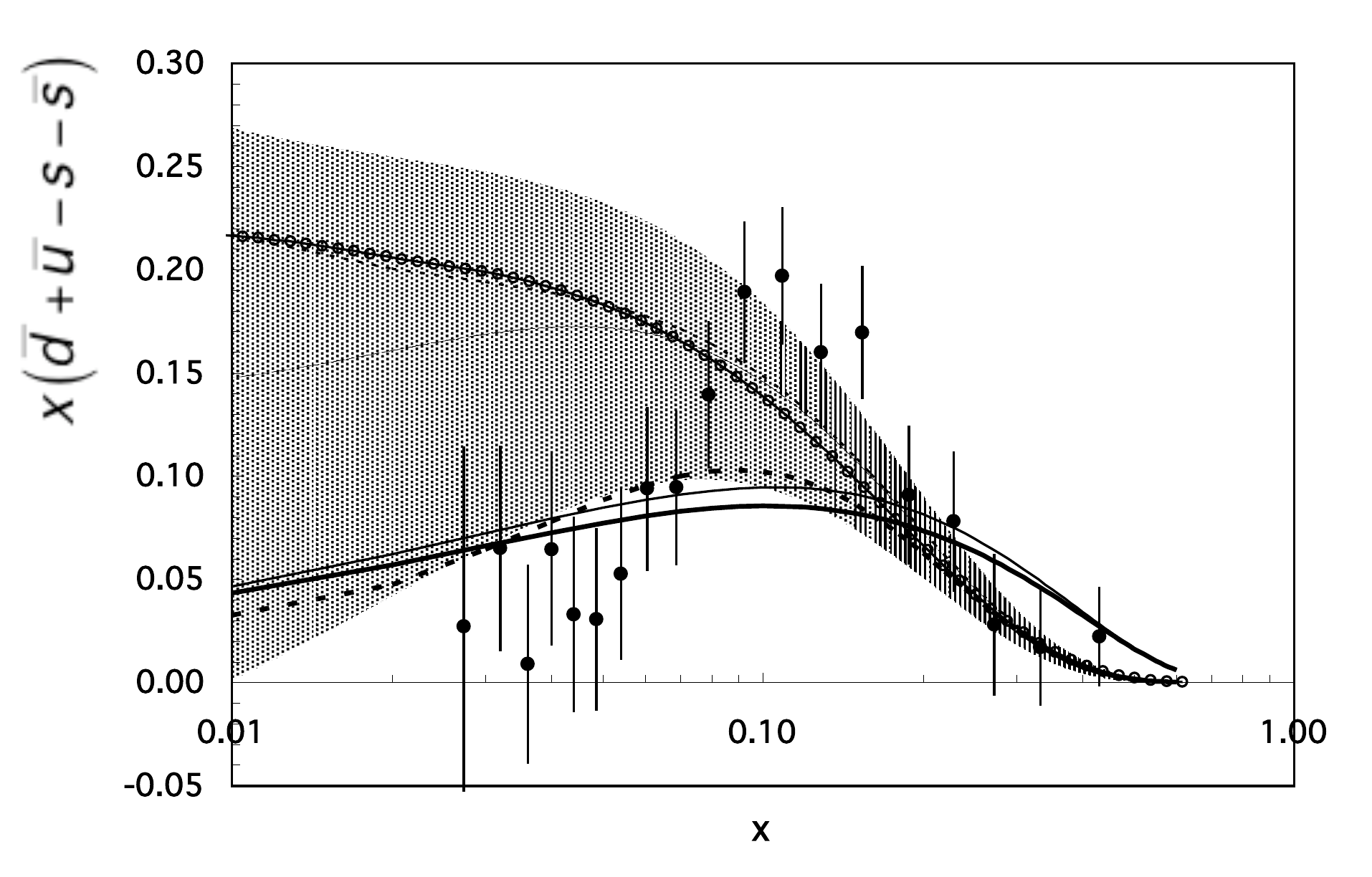}
\caption{Left: MCM calculations for SU(2) flavor asymmetry in the polarized nucleon sea, compared with HERMES measurements \cite{HERMES04}.
Right: 
MCM calculations for $x\Delta(x)$ (thick solid curve - with $K^*$ contributions; thin solid curve - without $K^*$ contributions), compared 
with the `experimental' distributions based on HERMES \cite{HERMES04} (data points)  and NuTeV \cite{MasonNuTeVNLO} (thin solid curve with open-circle markers) data,
and the results from CTEQ6.6M \cite{CTEQ6.6} (thick dashed curve), MSTW2008 \cite{MSTW2008} (thin dashed curve)
and CTEQ6.5 \cite{CTEQ6.5S0} (shaded area with the middle curve giving the central values).}
\label{fig_SU2+3}
\end{figure}

The numerical results for 
$x(\Delta \ubar -\Delta \dbar)$ are shown in Fig.~\ref{fig_SU2+3} together with HERMES data \cite{HERMES04}.
Our theoretical calculations are consistent with the data, although large uncertainties exist in the data.
Also we found that the SU(2) flavor symmetry breaking in the polarized nucleon sea is much smaller
than that in the unpolarized sea, which is in contrast to calculations using chiral quark soliton
model which predicts the differences $\Delta \dbar - \Delta \ubar$
and $\dbar - \ubar$ are similar in magnitude \cite{Chiral}.

There may be other non-perturbative contributions to flavor symmetry breaking of
the parton distributions of the bare nucleon.
Some studies \cite{FCaoS_plzedEPJC,PauliBlocking}
estimated that these contributions could be significantly larger than the contributions
from the meson cloud by considering Pauli blocking effects.
However the HERMES data indicate that these
non-perturbative contributions from the bare nucleon cannot be very large.
As Pauli blocking effects are expected to be of similar size in both polarized and unpolarized
cases, this conclusion may be of important in discussions of
$\dbar - \ubar$ difference \cite{MST98}.

The results for the SU(3) flavor symmetry breaking in the unpolarized nucleon sea, defined as
\bea
\Delta (x)=\dbar(x)+\ubar(x)-s(x)-\sbar(x),
\label{Delta}
\eea
are shown in Fig.~\ref{fig_SU2+3}. 
The largest contributions to the $\dbar$ and $\ubar$ sea come from the Fock states $\left | N \pi \right >, \left | N \rho \right >$ and 
$ \left | \Delta \pi \right >$, while the Fock states involving $K$ and $K^*$ mesons and responsible for the strange sea
are of roughly equal magnitude.
We also show the results when Fock states containing $K^{*}$ mesons are omitted in order to provide some estimate of the uncertainties in the model calculation. 
We can see that this has a small effect on the calculation of $x\Delta(x)$, and gives us some confidence that the uncertainties in the MCM calculations are under control.

The results for $x\Delta(x)$ are in reasonably good agreement with HERMES results \cite{HERMES08} except for the 
region around $x \sim 0.10$, but smaller than the next-leading-order analysis \cite{MasonNuTeVNLO} of NuTeV data \cite{NuTeV}.
We note that the two `experimental' distributions which are obtained by using HERMES \cite{HERMES08} and NuTeV \cite{MasonNuTeVNLO} data for the 
$s+\sbar$ together with CTEQ parameterizations \cite{CTEQ6.5S0,CTEQ6.6} of $x(\dbar+\ubar)$ do not agree well. 
It would be very interesting to see if this difference remains when the HERMES data are analyzed at next-to-leading order (NLO) in QCD. 
The MCM calculations for $x\Delta(x)$ also agree with the distribution obtained with CTEQ6M \cite{CTEQ6.6} set for the PDFs,
but significantly smaller than that obtained with MSTW2008 \cite{MSTW2008} set for the PDFs
and the central values of CTEQ6.5 \cite{CTEQ6.5S0} set for the PDFs. 
It is worth noting that the MCM calculations of $x\Delta(x)$ are independent of any global PDF set for the proton.

\subsection{Quark-antiquark symmetry breaking}

There has been interest for some time in the question of whether 
non-perturbative processes can lead to a difference between the strange 
and antistrange quark distribution functions of the proton \cite{SignalT87}.
More recent interest in this topic was prompted by the 
measurement of $\sin^{2} \theta_{W}$ by the NuTeV collaboration \cite{NuTeV02}. 
The large difference between the NuTeV result and the accepted value 
of around three standard deviations could arise, or be partly explained by, 
a positive value of the second moment of the difference between the strange and antistrange 
distributions, $\langle x(s-\bar{s})\rangle=\int_{0}^{1} dx x[s(x) - \bar{s}(x)]$,
as has been pointed out by Davidson and co-workers \cite{Davidson02}.

\begin{figure}[htb]
\includegraphics[width=15pc]{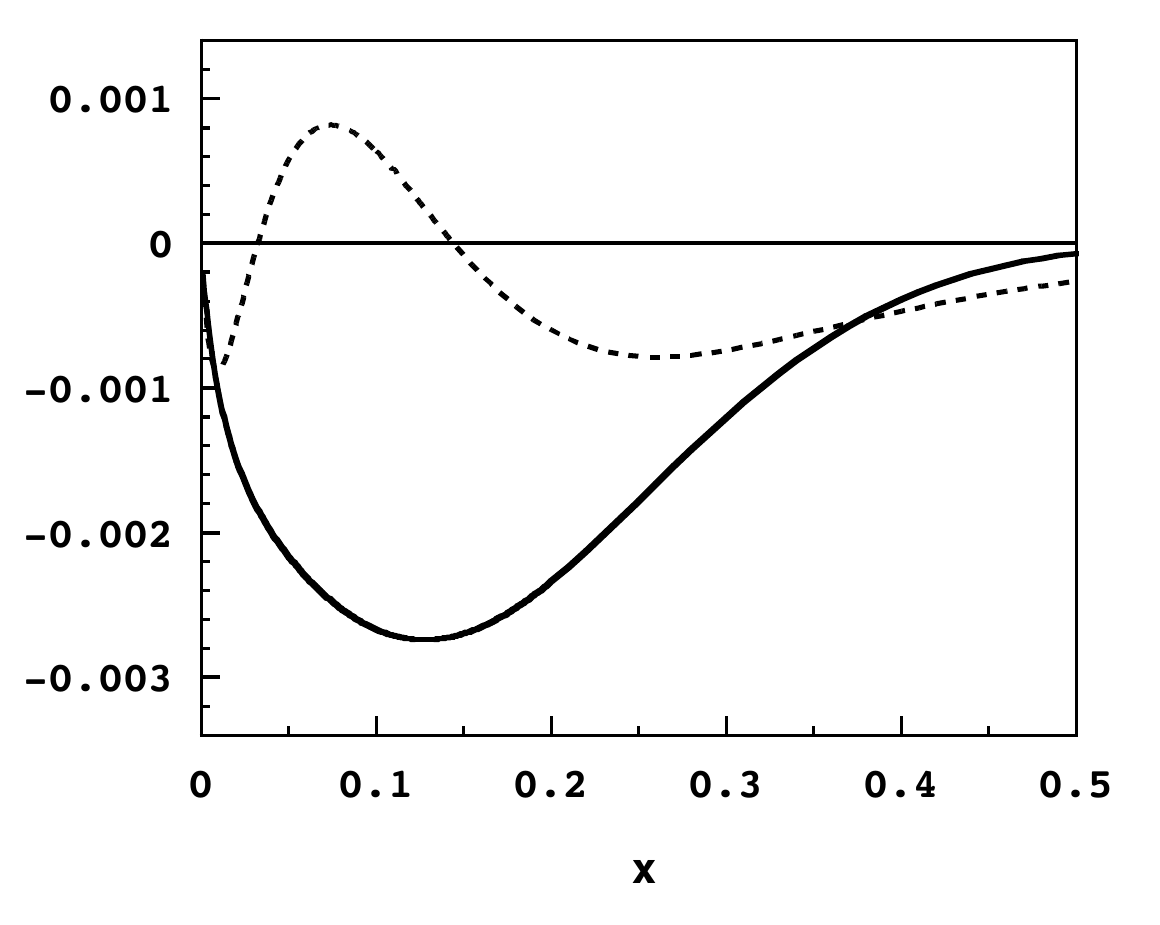} {\hskip 0.5cm}
\label{fig_ssbarDssbar}
\caption{The strange-antistrange symmetry breaking in the unpolarized (dashed curve for $x(s-\sbar)$ at $Q^2=16$ GeV$^2$)
and polarized (solid curve for $x(\Delta s - \Delta \sbar)$ at $Q^2=2.5$~GeV$^2$) nucleon sea.}
\end{figure}

We study the strange-antistrange asymmetry by considering
the fluctuations to $\Lambda K^{*}$ and $\Sigma K^{*}$ in addition to the 
$\Lambda K$ and $\Sigma K$ fluctuations of the proton. The results are shown in
Fig.~\ref{fig_ssbarDssbar} for the unplolarized and polarized PDFs.
It can be seen that  the strange-antistrange symmetry breaking is more significant in the polarized sea 
than in the unpolarized sea.

We calculate the second moment of the strange-antistrange asymmetry both with and 
without the $K^{*}$ contributions. 
It was found that 
$\langle x(s-\bar{s})\rangle=
1.43 \times 10^{-4}~(-1.35 \times 10^{-4})$ without (with) the $K^{*}$ states.
Omitting the fluctuations involving $K^{*}$ mesons changes the sign of 
the second moment of the asymmetry, so it is important that these fluctuations are 
included in any discussion of the asymmetry of the strange sea. 
The strange-antistrange asymmetry calculated in the MCM is an order of magnitude
too small to have any significant effect on the NuTeV result for the weak mixing angle. 
 
\subsection{Charge symmetry breaking}

It has been generally believed that charge symmetry is highly respected in the nucleon system.
However some theoretical calculations have suggested that the charge symmetry breaking (CSB)
in the valence quark distributions may be as large as $2\% \sim 10\%$ (see e.g. \cite{Sather,RodionovTL})
which is rather large compared with the low-energy results.
Any unexpected large CSB will greatly affect our understanding
of non-perturbative dynamics and hadronic structure,
and also the extraction of $\sin^2\theta_W$ from neutrino scattering \cite{Davidson02}.

At the quark level charge symmetry implies the invariance of a system
under the interchange of up and down quarks.
The breaking of the charge symmetry in the parton distributions of the nucleon
can be `measured' via the quantities:
\bea
\delta d_v(x)=d^p_v(x)-u^n_v(x),  & &
\delta u_v(x)=u^p_v(x)-d^n_v(x),   \nonumber \\
\delta \dbar(x)=\dbar^p(x)-\ubar^n(x), & &
\delta \ubar(x)=\ubar^p(x)-\dbar^n(x),  \\
\delta s(x)=s^p(x)-s^n(x), \nonumber & &
\delta \sbar(x)=\sbar^p(x)-\sbar^n(x). \nonumber
\label{CSV}
\eea

The charge symmetry breaking in both the valence and sea distributions has a non-perturbative origin.
Thus we can investigate the CSB using the meson cloud model.
The baryons (mesons) in the respective virtual Fock states of the proton
and neutron may differ in the charge they carry. If we neglect the mass differences between these baryons (mesons)
the fluctuation functions for the proton and neutron will be the same, and the contributions to the parton distributions of
the nucleon from these fluctuations will be charge symmetric.
As is well known, the electromagnetic interaction induces mass differences
among these baryons (mesons). If we take into account these mass
differences the probabilities for the corresponding fluctuations of proton and neutron will be different, and
thus the contributions to the parton distributions of the proton and neutron will be different, which results in CSB
in the parton distributions of the nucleon. Although it is argued from the quark model calculations
that the electromagnetic effect does not play a significant role in the calculation of CSB in the parton distributions \cite{RodionovTL},
it is worthwhile to study this effect using a different theoretical picture.

\begin{figure}[htb]
\includegraphics[width=15pc]{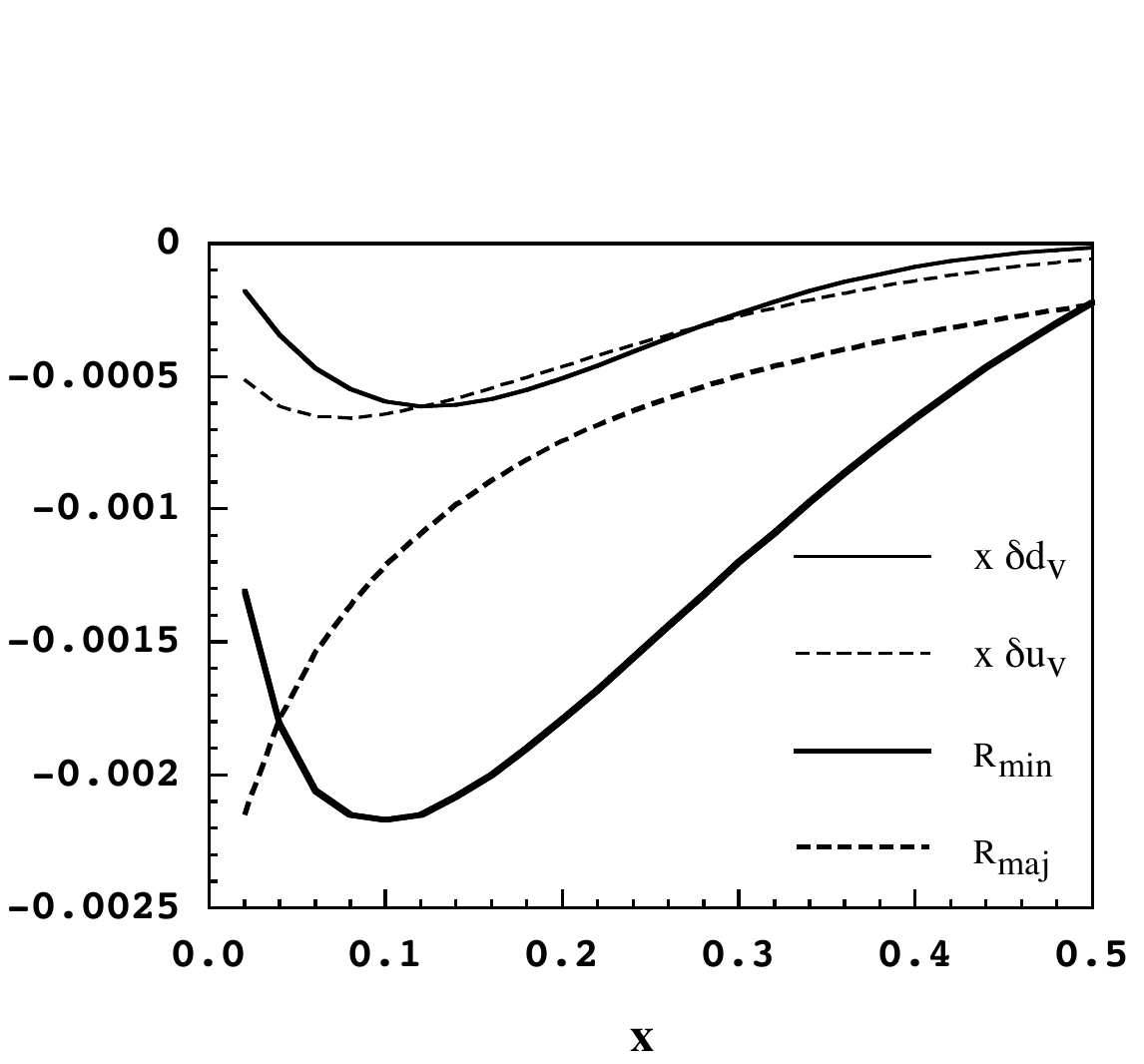}{\hskip 0.5cm}
\includegraphics[width=15pc]{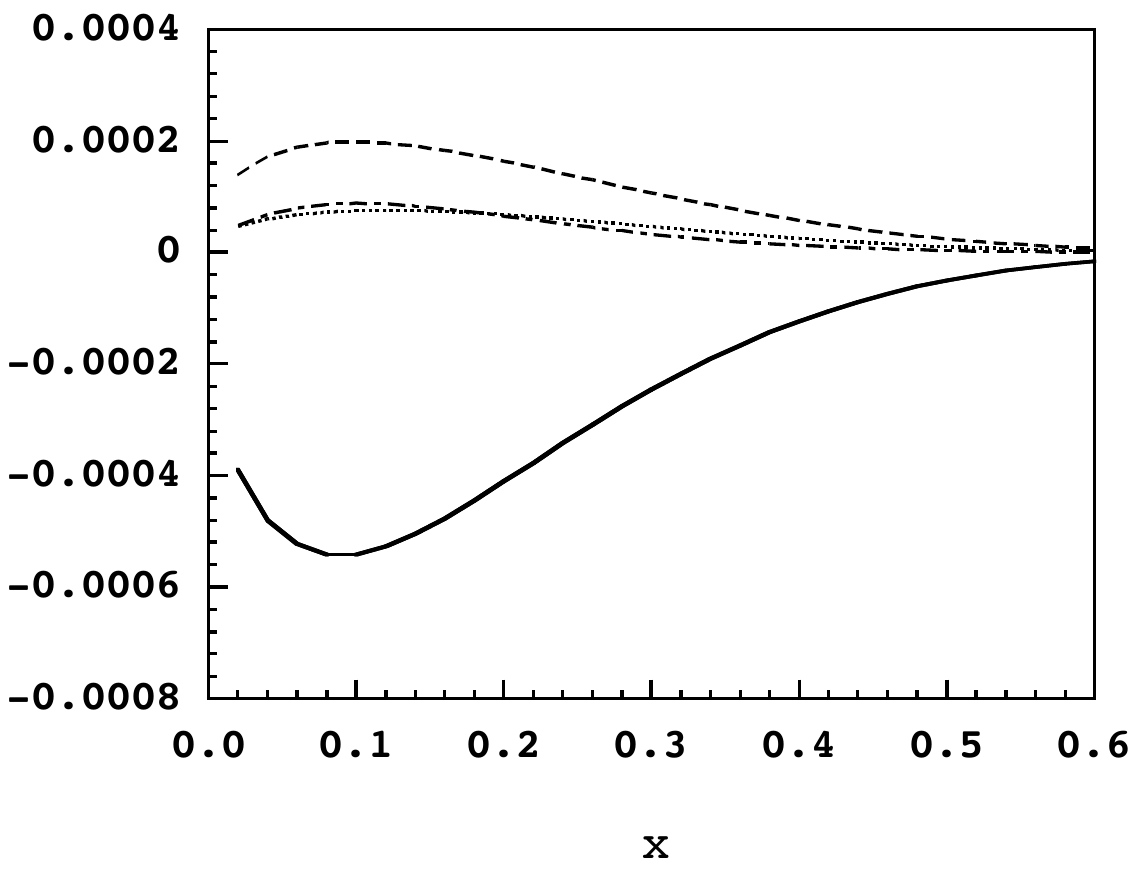}{\hskip 0.5cm}
\caption{Left: The charge symmetry breaking in the valence quarks.
$R_{\rm min}=\delta d_v/d^p_v$ and
$R_{\rm maj}=\delta u_v/u^p_v$. $Q^2=4$ GeV$^2$.
Right: The charge symmetry breaking in the sea quarks,
$x \delta \dbar$ (solid curve), $x \delta \ubar$ (dashed curve),
$x \delta s$ (dash-dotted curve), and $x \delta \sbar$ (dotted curve).
$Q^2=4$ GeV$^2$.}
\label{fig_CSB}
\end{figure}

The results for the CSB in the valence quarks
($x \delta d_v$ and $x \delta u_v$) and sea quarks ($x\delta \dbar$, $x\delta \ubar$, $x \delta s$ and $x \delta \sbar$)
are shown in Fig. \ref{fig_CSB}.
We find that $x \delta d_v$ and $x \delta u_v$ have similar shape and both are negative, which is quite different from the
quark model prediction of $x \delta d_v$ being positive for most values of $x$ \cite{Sather,RodionovTL}.
Furthermore, our numerical results are about 10\% of the quark model estimations.
It has been argued that although the absolute values of $\delta d_v$ and 
$\delta u_v$ may be small, the ratio $R_{\rm min}=\delta d_v/d^p_v$ can 
be much larger than the ratio $R_{\rm maj}=\delta u_v/u^p_v$ in the 
large-$x$ region since $d^p_v(x)/u^p_v(x) \ll 1/2$ as $x \ra 1$, and values 
as large as $10\%$ \cite{RodionovTL} have been obtained for the ratio $\delta d_v/d^p_v$.
No such large-$x$ enhancement appears in our calculation for both ratios.
The smallness of any CSB effect as $x \ra 1$ is natural in the MCM,
as all the fluctuation functions go to zero as $y \ra 1$,
and hence there is no non-perturbative contribution to the parton distributions at large $x$.
While quark models naturally predict significant CSB in the parton distributions
at large $x$, this is very hard to test, as all present-day experimental data
in this region has large uncertainties -- for instance the value of the ratio
$d/u$ as $x \ra 1$ is not well determined.
We did not find any significantly large CSB in the sea quark distributions of the
nucleon, which is consistent with recent phenomenological analysis \cite{BorosLT}.

\section{The total strange and antistrange quark distribution in the nucleon}

The non-strange light quark sea distributions are well determined by the global PDF fits to all available experimental data.  
Combining the global fit results for $\dbar(x)+\ubar(x)$ with our calculations for  $\Delta(x)$ we are able to estimate
the total strange sea distribution,
\bea
x\left[ s(x)+\sbar(x)\right]= x\left[ \dbar(x)+\ubar(x)\right]_{\mbox{Fit}} - x\Delta(x).
\label{xS+}
\eea

\begin{figure}
\includegraphics[width=18pc]{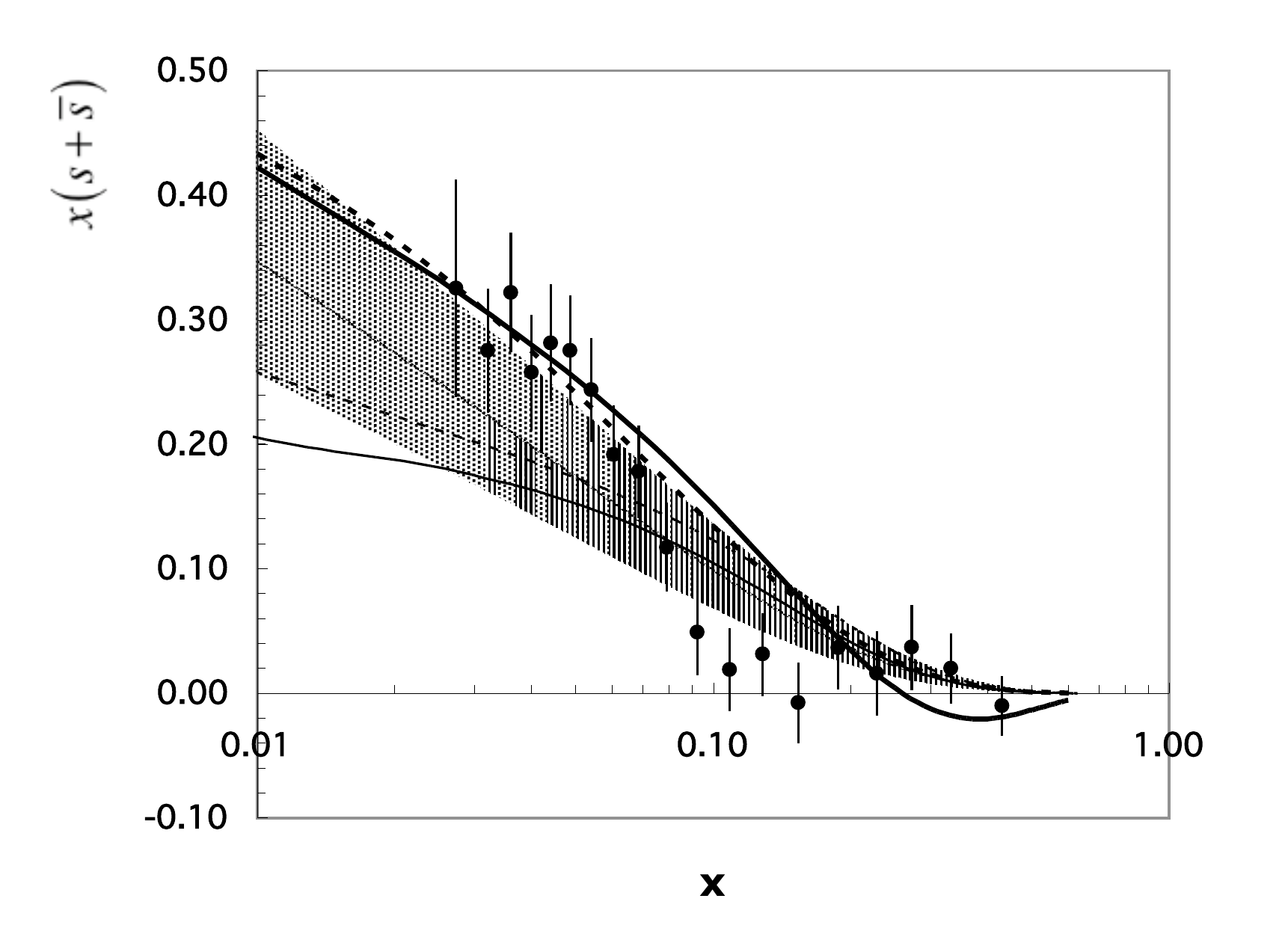} 
\label{fig_xsPsbar}
\caption{
The sum of the strange and antistrange quark distributions, shown as $x(s+\sbar)$, from the MCM calculations (thick solid curve),
 the HERMES measurement \cite{HERMES08} (data points), and 
 the next-to-leading order analysis of NuTeV dimuon data \cite{MasonNuTeVNLO} (thin solid curve).
 The global fit results from CTEQ6.6M \cite{CTEQ6.6} (thick dashed curve), 
 MSTW2008 \cite{MSTW2008} (thin dashed curve) and CTEQ6.5 \cite{CTEQ6.5S0} (shaded area) are also shown.}
\end{figure}

The results for $x(s+\sbar)$ are shown in Fig.~\ref{fig_xsPsbar}.
We have combined our calculations of $x\Delta(x)$ with the $(\dbar +\ubar)(x)$ distribution from the CTEQ6.6M set for the PDFs of the nucleon.
We find that our calculations are in good agreement with the CTEQ6.6M PDF set \cite{CTEQ6.6}, and with the HERMES data \cite{HERMES08}
for the region $x < 0.07$,
but are larger than those from the MSTW2008 \cite{MSTW2008} and CTEQ6.5 \cite{CTEQ6.5S0} PDF sets,
and the NLO analysis of NuTeV data \cite{MasonNuTeVNLO}.
Our calculations for $x(s+\sbar)$ becomes negative for $x > 0.25$ which is unreasonable. 
The reason for this could be that our model calculations overestimate $x\Delta(x)$, or that
$x(\dbar+\ubar)(x)$ is underestimated in the CTEQ6.6 PDF set, or both.
We also note that omitting the $K^{*}$ states from the calculation makes almost no difference in this region.
Our calculations suggest that the suppression of strange sea relative to the 
non-strange light antiquark sea is not as large as has been previously believed.

\section{Summary}

The possible breaking of various parton model symmetries by the nucleon's parton
distribution functions is of great interest since it can provide important information for the non-perturbative structure
of hadrons and the strong interaction. We investigate the breaking of flavor symmetry, quark-antiquark symmetry and charge symmetry
in the unpolarized and polarized nucleons using the meson cloud model.
We find that the SU(2) flavor symmetry in the polarized nucleon sea is broken
($\Delta \ubar < \Delta \dbar$), and the symmetry breaking is much smaller than that in
the unpolarized nucleon sea.
The strange-antistrange symmetry is found to be
broken in both the unpolarized and polarized nucleon sea,
and the asymmetry is more significant in the polarized sea than that in the unpolarized sea.
The charge symmetry breaking is found to be small.
We estimated the total strange and antistrange quark distribution of the nucleon and found
that the suppression of strange sea relative to the 
non-strange light antiquark sea is not as large as has been previously believed.

\begin{theacknowledgments}
I would like to thank Tony Signal and  Francois Bissey for many years of enjoyable collaboration on the study of nucleon structure.
\end{theacknowledgments}

%%%%%%%%%%%%%%%%%%%%%%%%%%%%%%%%%%%%%%%%%%%%
%% Sample figure:
%%
%% The option [height=...] scales the picture to the given height,
%% without it it would be printed at its nominal size
%%%%%%%%%%%%%%%%%%%%%%%%%%%%%%%%%%%%%%%%%%%%

%\begin{figure}
%  \includegraphics[height=.3\textheight]{golfer.ps}
% \caption{Picture to fixed height}
%\end{figure}
%%%%%%%%%%%

\bibliographystyle{aipproc}   % if natbib is available
%\bibliographystyle{aipprocl} % if natbib is missing

%%%%%%%%%%%%%%%%%%%%%%%%%%%%%%%%%%%%%%%%%%%
%% You probably want to use your own bibtex database here
%%%%%%%%%%%%%%%%%%%%%%%%%%%%%%%%%%%%%%%%%%%
\bibliography{sample}

%%%%%%%%%%%%%%%%%%%%%%%%%%%%%%%%%%%%%%%%%%%
%% Just a reminder that you may have to run bibtex
%% All of it up to \end{document} can be removed
%% if you don't like the warning.
%%%%%%%%%%%%%%%%%%%%%%%%%%%%%%%%%%%%%%%%%%%
\IfFileExists{\jobname.bbl}{}
 {\typeout{}
  \typeout{******************************************}
  \typeout{** Please run "bibtex \jobname" to optain}
  \typeout{** the bibliography and then re-run LaTeX}
  \typeout{** twice to fix the references!}
  \typeout{******************************************}
  \typeout{}
 }

%%%%%%%%%%%%%%%%%%%%%%%%%%%%%%%%%%%%%%%%%%%
%% The following lines show an example how to produce a bibliography
%% without the help of the BibTeX program. This could be used instead
%% of the above.
%%%%%%%%%%%%%%%%%%%%%%%%%%%%%%%%%%%%%%%%%%%

\end{document}